\documentclass[preprint,aps,prd,groupedaddress,showpacs,showkeys]
              {revtex4}%
\usepackage{graphicx}
\usepackage{amsmath}
\usepackage{bm}
\usepackage{relsize}
\RequirePackage{xspace}
\begin{document}

\title{
Next-to-leading order Calculation of a Fragmentation Function
\\ in a Light-Cone Gauge} 

\author{Jungil Lee}
\affiliation{Department of Physics, Korea University, Seoul 136-701, Korea}



\date{\today}
\begin{abstract}
The short-distance coefficients for the color-octet
$^3S_1$ term in the fragmentation function for a gluon to split into
polarized heavy quarkonium states are re-calculated to order $\alpha_s^2$.
The light-cone gauge remarkably simplifies the calculation by eliminating many 
Feynman diagrams at the expense of introducing spurious poles in loop 
integrals. We do not use any conventional prescriptions for spurious pole. 
Instead,
we only use gauge invariance with the aid of Collins-Soper definition of
the fragmentation function. 
Our result agrees with a previous calculation
of Braaten and Lee in the Feynman gauge, but disagrees with another previous
calculation.
\end{abstract}

\pacs{13.87.Fh, 13.60.Le, 12.38.-t, 12.38.Bx, 13.88.+e}
\keywords{Fragmentation, Fragmentation function, Next-to-leading order,
           Light-cone gauge, Quarkonium}

\maketitle

\section{introduction}
Heavy quarkonium state is one of the simplest hadron with which we can probe 
both perturbative and nonperturbative nature of quantum chromodynamics.
Among various quarkonia, $S$-wave spin-triplet states are especially
interesting because of their clean leptonic decay modes. 
Based on factorization theorems for inclusive single-hadron 
production~\cite{Collins:1987pm,C-S-book},
one can deduce that dominant production mechanism for heavy quarkonia with 
large transverse momentum $p_T$ is \textit{fragmentation}~\cite{FM}, 
the production of a parton 
which subsequently decays into the quarkonium state and other partons.
In the region, where fragmentation dominates, theoretical calculation becomes
simplified and more reliable. Measurements of large-$p_T$ $S$-wave 
spin-triplet charmonia production cross section at the Fermilab Tevatron has 
led to a remarkable progress in heavy quarkonium physics based on 
nonrelativistic QCD~(NRQCD)~\cite{Bodwin:1994jh}.
In high-energy $p \bar p$ collisions, gluon production rate is dominant
and inclusive production of a heavy quark pair
$Q \overline{Q}$ via subsequent decay of this almost on-shell gluon
is enhanced by the gluon propagator~\cite{B-Y:S1}.
Unexpectedly large measured production rate of direct $J/\psi$ and $\psi'$ 
at large $p_T$ at the Tevatron~\cite{Abe:1997jz} was explained 
by the gluon fragmenation into a color-octet $Q \overline{Q}$ pair 
followed by a nonperturbative NRQCD transition into the spin-triplet $S$-wave 
quarkonia~\cite{B-F}. Importance of the gluon fragmentation mechanism has
been tested also in inclusive $J/\psi$ production in $Z^0$ decay~\cite{Z0}. 
There are still two
open problems in the field. One is the polarization of prompt  $J/\psi$
at the Tevatron~\cite{CDF-pol,pol}. 
The other is the cross section for exclusive 
$e^+e^-\to J/\psi+\eta_c$~~\cite{Abe:2002rb,B-th} and  
$e^+e^- \to J/\psi+c\bar{c}+X$ at $B$-factories~\cite{Abe:2001za,Jpsi-X}.
However, the color-octet mechanism in NRQCD has been tested successfully in 
various ways. 
Comprehensive reviews on NRQCD phenomenology can be found in 
Ref.~\cite{REV}. In the rest of this paper, we shall restrict our discussion 
to the gluon fragmentation into a color-octet spin-triplet $Q\overline{Q}$ 
pair evolving into a $S$-wave heavy quarkonium.

In the NRQCD factorization formalism~\cite{Bodwin:1994jh},
the fragmentation function $D(z,\mu)$ for a parton splitting
a heavy quarkonium is expressed as a linear combination of NRQCD matrix
elements, which can be regarded as phenomenological parameters.
Here, $z$ is the momentum fraction of the final hadron relative to
the decaying parton and $\mu$ is the hard-scattering scale of the process.
Corresponding short-distance factors depend on $z$ and are calculable
in perturbation theory.
Most of the phenomenologically relevant short-distance factors
have been calculated to leading order in $\alpha_s$.
They all begin at order $\alpha_s^2$ or higher, with the exception
of the color-octet $^3S_1$ term in the gluon fragmentation function,
which begins at order $\alpha_s$~\cite{B-F}. The color-singlet $^3S_1$ 
channel is suppressed because the short-distance factor 
begins at order $\alpha_s^3$~\cite{B-Y:S1}.
Since  the color-octet $^3S_1$ term dominates, the high-$p_T$
gluon fragmentation phenomena in heavy quarkonium production,
the next-to-leading order correction of order $\alpha_s^2$
to this term is particularly important.
Unfortunately, two available results for the color-octet
$^3S_1$ term disagree with each other~\cite{Ma-2,Braaten:2000pc}.
Therefore, it is worth while to calculate this important function
in an independent way.
Since both previous calculations employed the Feynman gauge,
we shall present our results in the light-cone gauge.

The light-cone gauge is a physical gauge where the gluon field $A^\mu$ has 
vanishing light-cone projection $A\cdot n=0$, where $n$ is an arbitrary light-like($n^2=0$)
vector appearing in the gauge-fixing term in the QCD Lagrangian.
Derivation of the Altarelli-Parisi evolution of parton 
densities~\cite{Altarelli:1977zs}
is one of the best examples of the use of the light-cone gauge.
In a light-cone gauge,
in which the fragmentation function was originally defined \cite{C-F-P},
there is a great simplification in QCD calculations.
The eikonal line as well as the ghost decouples from the gluon, since the
coupling, proportional to $n^\mu$, is orthogonal to the gluon
propagator.
However, one draw-back in higher-order calculations is existence of 
the spurious pole $1/k\cdot n$ in the gluon propagator
\begin{eqnarray}
\frac{i}{k^2+i\epsilon}\left[
-g^{\mu\nu}+\frac{k^\mu n^\nu+n^\mu k^\nu}{k\cdot n}
\right],\; \epsilon\to 0^+,
\label{Gpro:LC}
\end{eqnarray}
where $k$ is the momentum of the gluon.
One should be very careful in dealing with the pole in evaluating loop integrals.
As conventional methods, Cauchy-Principal-Value~(CPV) or 
Mandelstam-Leibbrandt~(ML) 
prescriptions~\cite{Mandelstam:1982cb,Leibbrandt:1983pj} 
have been used for a long time.
The two prescriptions follow from different canonical 
quantizations,  light-front~(LF) and equal-time~(ET), respectively~\cite{conv}.
It turns out that LF quantization~(CPV) is untenable beyond the tree level
as it conflicts with causality: the resulting theory cannot be renormalized.
The ML prescription is causal; renormalization has been proved not only
for the Green functions at any perturbative order~\cite{Bassetto:1987sw}
but also for composite operators~\cite{Acerbi:1993xk}.
All previous calculations beyond-leading order had to employ some
prescription for the spurious pole.
The CPV prescription is used in Refs. \cite{C-F-P,C-F-P-2,AP-2-PV}.
The ML prescription is  employed in the leading order \cite{AP-1} and
the next-to-leading order \cite{AP-2-1,AP-2-2}.
Comprehensive reviews on these prescriptions can be found in 
Refs.~\cite{Leibbrandt:1987qv,BOOK1,BOOK2}.

In our light-cone gauge calculation, 
we introduce a new method in determining the spurious pole 
appearing in our calculation for the fragmentation function. 
We first employ a light-cone gauge to express the fragmenation function
in terms of one-loop scalar integrals using the gauge-invariant definition 
of Collins and Soper~\cite{C-S}.  
However, we do not impose the sign of $i\epsilon$ 
in the spurious poles and keep the scalar integrals from evaluation
unlike the conventional ways.
Our light-cone-gauge result for the fragmenation function
is same as that for the Feynman gauge, which is guaranteed by the
gauge-invariant Collins--Soper definition. Therefore, the scalar integrals
involving spurious poles can be determined by comparing with the Feynman-gauge 
result. The spurious poles are finally identified with poles coming 
from the eikonal-line contribution in the Feynman gauge. Because all the 
poles are well-defined in the Feynman gauge, the spurious-pole contributions 
in the light-cone gauge are completely determined based on gauge invariance.
Our result agrees with a previous calculation of 
Braaten and Lee~\cite{Braaten:2000pc} in the Feynman gauge, 
but disagrees with the other previous calculation~\cite{Ma-2}.

\section{Collins-Soper definition and light-cone gauge}
The fragmentation function $D_{g \to H}(z,\mu)$
gives the probability that a gluon produced in a hard-scattering process
involving momentum transfer of order $\mu$ decays into a hadron $H$
carrying a fraction $z$ of the gluon's longitudinal momentum.
This function can be defined in terms of the matrix element of a bilocal 
operator involving two gluon field strengths in a light-cone 
gauge~\cite{C-F-P}.  In Ref.~\cite{C-S},
Collins and Soper introduced a gauge-invariant definition of the gluon 
fragmentation function that involves the matrix element 
of a nonlocal operator consisting of two gluon field strengths and 
eikonal operators.  One advantage of this definition is that it 
avoids subtleties associated with products of singular distributions.  
The gauge-invariant definition is also advantageous for explicit 
perturbative calculations, because it allows the calculation of 
radiative corrections to be simplified by using the Feynman gauge.   

The gauge-invariant definition 
of Collins and Soper for the gluon fragmentation function for
splitting into a hadron $H$ is~\cite{C-S}  
\begin{eqnarray}
D_{g \to H}(z,\mu) &=&
\frac{(-g_{\mu \nu})z^{N-2}}{16\pi(N-1) k^+}
\int_{-\infty}^{+\infty} dx^- e^{-i k^+ x^-}
\nonumber\\
&&\times
\langle 0 | G^{+\mu}_c(0)
{\cal E}^\dagger(0^-)_{cb} \; {\cal P}_{H(z k^+,0_\perp)} \;
{\cal E}(x^-)_{ba} G^{+ \nu}_a(0^+,x^-,0_\perp) | 0 \rangle\;.
\label{D-def}
\end{eqnarray}
The operator ${\cal E}(x^-)$ in (\ref{D-def}) is an eikonal operator
that involves a path-ordered exponential of gluon field operators along
a light-like path:
\begin{equation}
{\cal E}(x^-)_{ba} \;=\; {\rm P} \exp
\left[ +i g \int_{x^-}^\infty dz^- A^+(0^+,z^-,0_\perp) \right]_{ba},
\label{E}
\end{equation}
where $A^\mu(x)$ is the matrix-valued gluon field in the adjoint
representation:  $[ A^\mu(x) ]_{ac} = if^{abc} A_b^\mu(x)$.
The operator ${\cal P}_{H(p^+,p_\perp)}$ in Eq.~(\ref{D-def})
is a projection onto states that, in the asymptotic future, contain 
a hadron $H$ with momentum $p = (p^+,p^-=(m_H^2+ p_\perp^2)/p^+,p_\perp)$, 
where $m_H$ is the mass of the hadron.
The hard-scattering scale $\mu$ in Eq.~(\ref{D-def})
can be identified with the renormalization scale of the nonlocal operator.
The prefactor in the definition (\ref{D-def})
has, therefore, been expressed as a function of the number 
of spatial dimensions $N=3-2\epsilon$.
This definition is particularly useful when we 
use dimensional regularization to regularize ultraviolet divergences.  
If the production process of the hadron $H$ can be described by perturbation
theory, one can use the definition (\ref{D-def}) to calculate the fragmentation
function $D_{g\to H}(z,\mu)$ as a power series in $\alpha_s$.
In Ref.~\cite{C-S}, complete sets of Feynman rules for this perturbative
expansion for quark and gluon fragmentation functions are given.
Since Ma first used the definition~\cite{Ma-1}, the method became popular 
in quarkonium physics~\cite{Braaten:2000pc,Braaten:2001sz,Bodwin:2003wh}
 because of various reasons listed above.
By inserting the eikonal operator (\ref{E}), the operator consisting of two 
gluon fields with different locations becomes gauge invariant.
At higher order in $\alpha_s$, there are numerous diagrams
which have gluons coupled to the eikonal lines.
In the light-cone gauge, the contribution of the eikonal operator disappears
since the gluon decouples from the eikonal line.
Therefore, there is a great reduction in the number of Feynman diagrams.
On the other hand, the spurious pole contribution of the gluon propagator
appears in the light-cone gauge. 
However, the gauge invariance of this definition (\ref{D-def})
provides the  gauge transformation of the eikonal line contribution in the 
Feynman gauge into the spurious pole contribution in the light-cone gauge.
By comparing the final results for the gauge-invariant quantity  
$D_{g\to H}(z,\mu)$ from the two gauges, the spurious pole coming from 
the gluon propagator in the light-cone gauge can be fixed unambiguously.

\section{perturbative calculation}
In this section we perform the next-to-leading-order calculation for
the fragmenation function for a gluon splitting into a color-octet 
spin-triplet $Q\overline{Q}$ pair of order $\alpha_s^2$ in a light-cone gauge.
Although we carry out the calculation using a different gauge,
we use the same conventions as those presented in Ref.~\cite{Braaten:2000pc}. 
We do not reproduce the description on the theoretical background of 
the fragmentation function for heavy quarkonium production in NRQCD 
factorization formalism which is well explained in 
Ref.~\cite{Braaten:2000pc}.
Based on the NRQCD factorization formalism~\cite{Bodwin:1994jh},
the fragmentation function is written in a factorized 
form~\cite{Braaten:2000pc}:
\begin{eqnarray}
D_{g \to H}(z) &=&
\left[ (N-1) d_T(z) + d_L(z)  \right]
\langle {\cal O}_8^H(^3S_1) \rangle,
\label{D-sum}
\end{eqnarray}
where $d_T$ and $d_L$ are the short-distance coefficients for
the transverse and longitudinal contributions and
$\langle {\cal O}_8^H(^3S_1) \rangle$ is the color-octet $^3S_1$
matrix element  defined in Ref.~\cite{Bodwin:1994jh}.

\begin{figure}
\includegraphics[width=7cm]{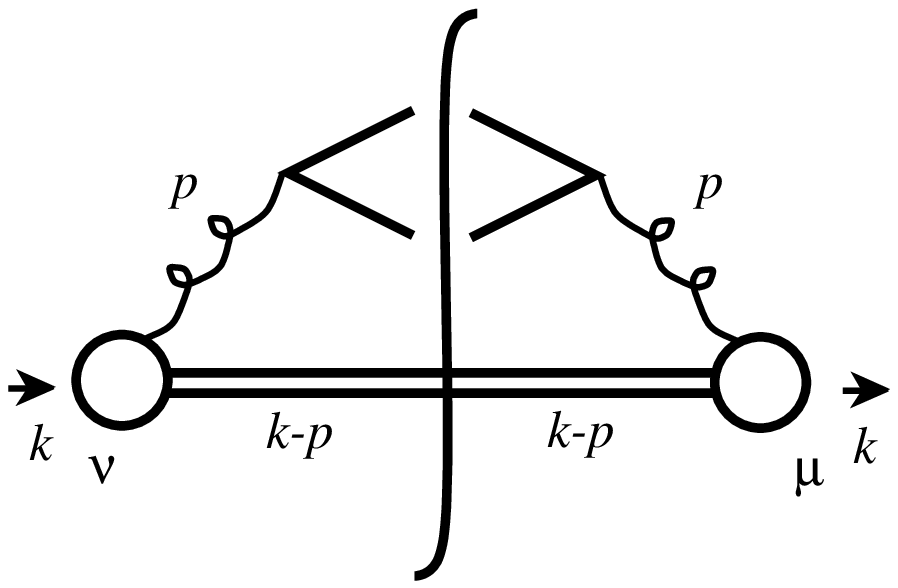}
\caption{\label{2body0}
Leading order Feynman diagram for $g \to Q\overline{Q}$.
}
\end{figure}

There is only one lowest-order diagram in both 
Feynman and light-cone gauge, which is shown in Fig.~\ref{2body0}.
The circles connected by the double pair of lines represent the
nonlocal operator consisting of the gluon field strengths and
the eikonal operators in the definition (\ref{D-def}).
The momentum $k = (k^+,k^-,k_\perp)$ flows into the circle on the left
and out of the circle on the right.
The cutting line represents the projection onto states which, in the
asymptotic future, include a $Q \overline{Q}$ pair with total momentum
$p = (z k^+, p^2/(z k^+), 0_\perp)$.
The appearance of the diagrams for both gauges is the same 
in this order, since the circle should emit a gluon.
With the Feynman rules of Ref.~\cite{C-S} and following the method of 
extracting the short-distance coefficients of the fragmentation function 
in Ref.~\cite{Braaten:2000pc},
we can read off the order-$\alpha_s$ terms in the short-distance functions
$d_T(z)$ and $d_L(z)$ as 
\begin{eqnarray}
d_T^{(\rm LO)}(z) &=&
\frac{\pi\alpha_s\mu^{2\epsilon}}{8N(N-1)m_Q^3}
\;\delta(1-z),
\label{d1-T}
\\
d_L^{(\rm LO)}(z) &=& 0.
\label{d1-L}\end{eqnarray}
We have neglected the relative momentum of the heavy quark 
in the $Q \overline{Q}$ rest frame so that the invariant mass of the
pair is $p^2 = 4 m_Q^2$.
The LO results (\ref{d1-T}) and (\ref{d1-L}) agree with previous calculations 
in the Feynman gauge~\cite{Braaten:2000pc,Braaten:1996rp}.
Note that there is no reason for worrying about the spurious pole in this
leading-order calculation.

\begin{figure}
\includegraphics[width=11.5cm]{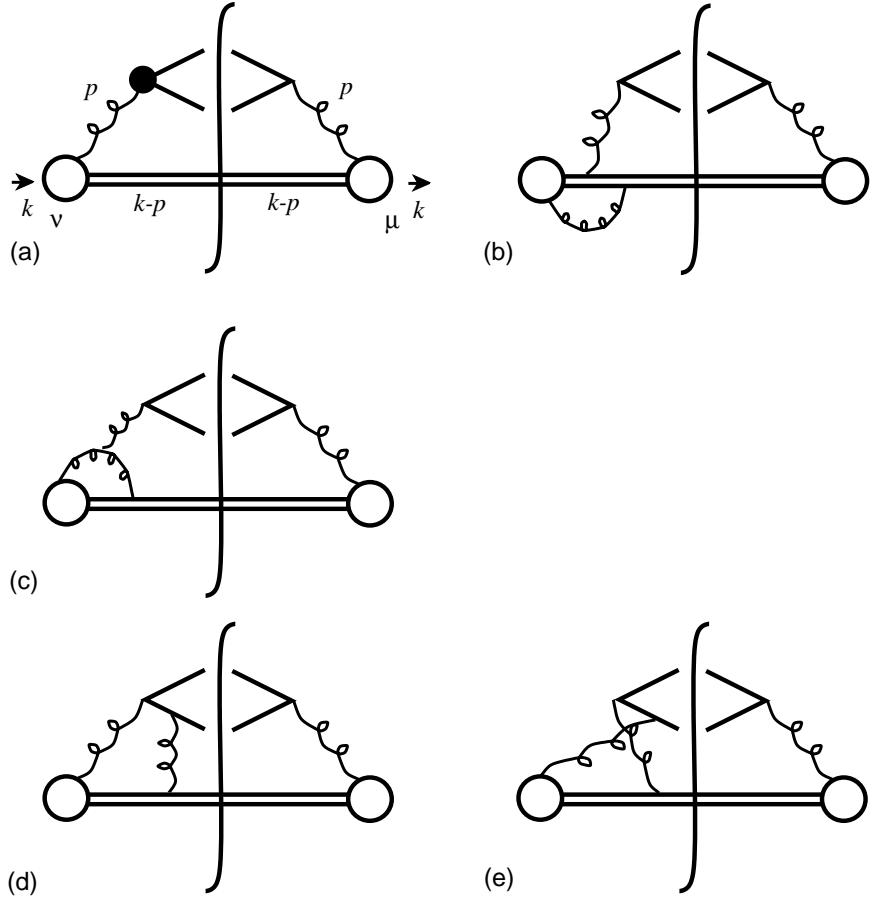}
\caption{\label{2bodya}
The Feynman diagrams of order $\alpha_s^2$ for $g \to Q\overline{Q}$
with $Q\overline{Q}$ final states.  There are additional contributions
from the complex-conjugate diagrams.
}
\end{figure}
\begin{figure}
\includegraphics[width=11.5cm]{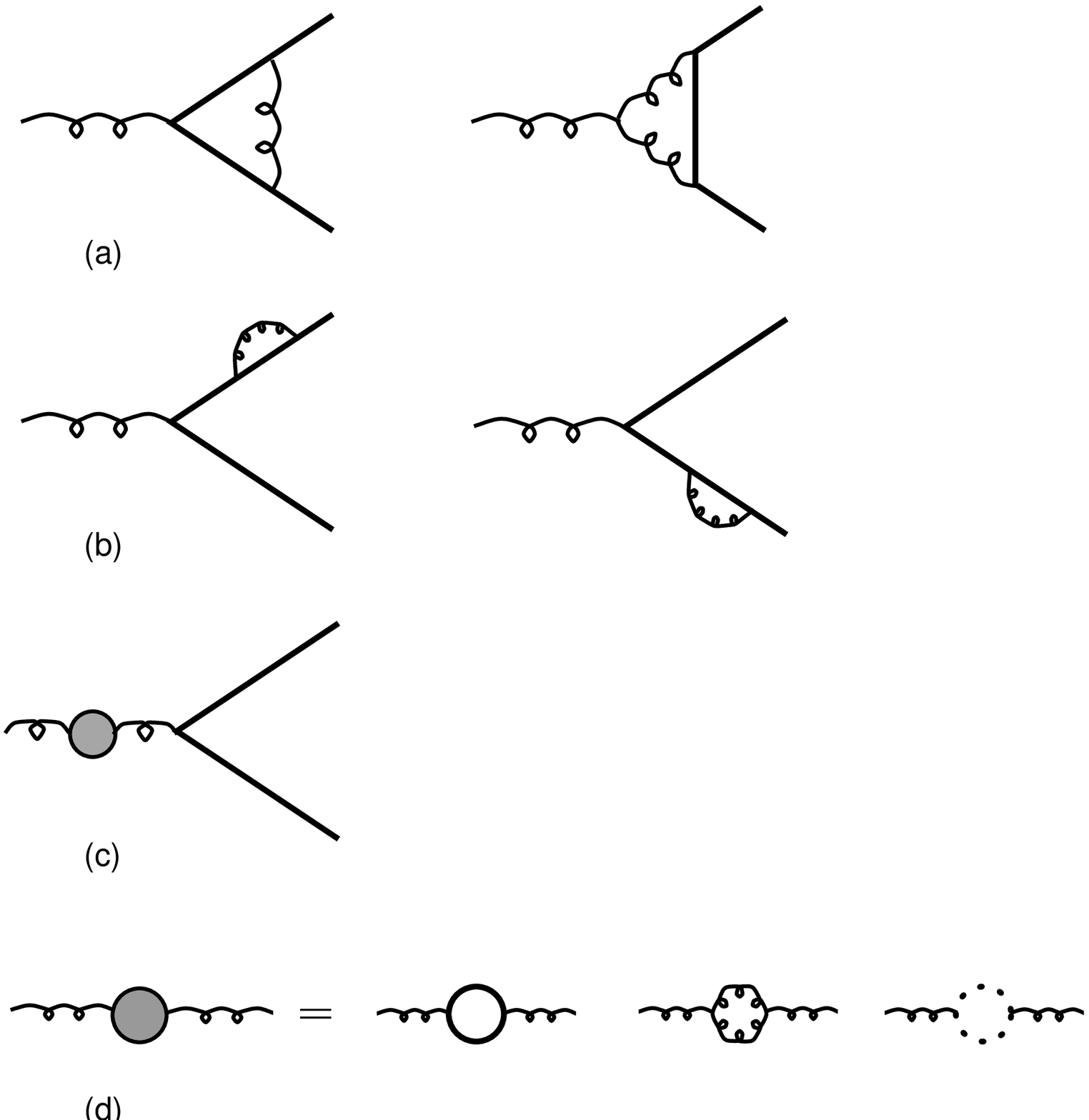}
\caption{\label{2onlya}
One loop correction diagrams for
$g^*\to Q\overline{Q}$.
}
\end{figure}

The Feynman diagrams for the 
fragmentation function for $g \to Q\overline{Q}$ at order $\alpha_s^2$
consist of virtual corrections, for which the final state is $Q\overline{Q}$,
and real-gluon corrections, for which the final state is $Q\overline{Q}g$.
The diagrams with virtual-gluon corrections to the left of the cutting line
are shown in Fig.~\ref{2bodya}.  The black blob in  Fig.~\ref{2bodya}(a)
includes the vertex corrections and propagator corrections shown in 
Fig.~\ref{2onlya}.
In the Feynman gauge, only the diagram in Fig.~\ref{2bodya}(b) vanishes,
because the gluon attached to the eikonal line gives a factor of $n^\mu$.
On the other hand, all the diagrams except for Fig.~\ref{2bodya}(a) 
vanish in the light-cone gauge.
If we use the threshold-expansion method of Braaten and 
Chen~\cite{Braaten:1996jt}, 
we can simplify the structure of the expression 
without employing the projection method. With the threshold expansion, 
we can keep the full structure of color and spin.  
Here we utilize the dimensionally regularized threshold expansion method 
of  Braaten and Chen~\cite{Braaten:1996rp,Braaten:1997rg}.
With the Dirac equation and the usual methods for reducing tensor integrals 
into scalar integrals, we factorize each virtual correction 
diagram into the leading order diagram in Fig.~\ref{2body0}, times 
a multiplicative factor.
In the light-cone gauge, the ghost decouples since its coupling to the
gluon is orthogonal to the gluon propagator (\ref{Gpro:LC}), 
so the gluon propagator correction factor shown in Fig. 3(d) does not 
have ghost contribution.

The virtual corrections contribute only to the transverse short-distance
function $d_T(z)$ defined in Ref.~~\cite{Braaten:2000pc}:
\begin{eqnarray}
d_T^{\rm (virtual)}(z) = d_T^{(\rm LO)}(z)\times
\;2\;{\rm Re}\bigg[\Lambda + \Pi + \delta Z_Q    + \Delta \bigg],
\end{eqnarray}
where $\Lambda$ is the vertex correction factor. The expressions 
$\delta Z_Q$ and $\Pi$ are given by
\begin{eqnarray}
\delta Z_Q^{\rm LC}&=&
i\frac{16\pi\alpha_s\mu^{2\epsilon}}{3}
\bigg[\left(2-N\right)I_{AD} + p^2 I_{ADD}
+ 2 p\cdot n I_{BCD}\bigg],
\label{Z:LC}
\\
\delta Z_Q^{\rm F}&=&
i\frac{16\pi\alpha_s\mu^{2\epsilon}}{3}
\bigg[\left(4-N\right)I_{AD} + p^2 I_{ADD} \bigg],
\label{Z:F}
\\
\Pi^{\rm LC}&=&
-i\;6\pi\alpha_s\mu^{2\epsilon}
\bigg\{
\bigg[7+\frac{1}{N} - \frac{2n_f}{3}\left(1-\frac{1}{N}\right)\bigg]\;I_{AB}
- 8p\cdot nI_{ABC}
\bigg\},
\label{Pi:LC}
\\
\Pi^{\rm F}&=&
-i\;6\pi\alpha_s\mu^{2\epsilon}
\bigg[3+\frac{1}{N} - \frac{2n_f}{3}\left(1-\frac{1}{N}\right)\bigg]I_{AB},
\label{Pi:F}
\end{eqnarray}
where the superscripts F and LC stand for Feynman gauge and light-cone
gauge, respectively.
The scalar integrals appearing in Eqs.(\ref{Z:LC})-(\ref{Pi:F}) are 
of the form
\begin{equation}
I_{AB\cdots} \;=\; \int\frac{d^{N+1}l}{(2\pi)^{N+1}} \frac{1}{AB\cdots}\;,
\end{equation}
where the denominator $AB\cdots$ can be a product of 1, 2, 3, or 4 of
the following factors:
\begin{eqnarray}
A&=&l^2+i\epsilon,\\
B&=&(l-p)^2+i\epsilon=l^2-2l\cdot p+4 m_Q^2+i\epsilon,\\
C&=&(p-l)\cdot n+i\epsilon,
\label{eq:C}
\\
D&=&(l-p/2)^2-m_Q^2+i\epsilon=l^2-l\cdot p+i\epsilon.
\end{eqnarray}
The values for the integrals are in the Appendix of 
Ref.~\cite{Braaten:2000pc} except for
\begin{eqnarray}
I_{AAD}&=&
\frac{-i}{(4\pi)^{2}(2m_Q^2)}
\left( \frac{4\pi }{ m_Q^2} \right)^\epsilon \;
\frac{\Gamma(1+\epsilon)}{\epsilon_{\rm IR}(1+2\epsilon)} \;,
\\
I_{ADD}&=&
\frac{i}{(4\pi)^{2}(2m_Q^2)}
\left( \frac{4\pi}{ m_Q^2} \right)^\epsilon \;
\frac{\Gamma(1+\epsilon) }{ \epsilon_{\rm IR}}.
\end{eqnarray}

Note that for the light-cone gauge integrals
$I_{BCD}$ in Eq.~(\ref{Z:LC}) and $I_{ABC}$ in Eq.~(\ref{Pi:LC}) 
have spurious poles $1/(p-l)\cdot n$ that require prescriptions.  
Those poles for the Feynman gauge
are unambiguously defined as in Eq.~(\ref{eq:C}).
Therefore, we postpone evaluating the scalar integrals for the time being.
If we use the ML prescription, each light-cone-dependent integral
has ultraviolet (UV) and IR structures
which are different from the values shown in the Appendix of 
Ref.~\cite{Braaten:2000pc}.
Effectively, the ML prescription transforms a double pole into an IR pole
and makes the integral satisfy naive power counting rules.
Note also that $Z_Q$ and $\Pi$ are gauge dependent.
The values in the Feynman gauge agree with well-known ones that can be
found, for example, in Ref.~\cite{Braaten:2000pc}.
The result using the ML prescription is known only for the UV poles.
We have full agreement in the UV poles if we use the ML prescription :
the gluon propagator correction term is proportional to the QCD beta function
as $\Pi=(33-2n_f)\alpha_s/(12\pi\epsilon_{\rm UV})$~\cite{Leibbrandt:1983pj,PI}
and $\delta Z^{\rm LC}_Q=\alpha_s/(3\pi\epsilon_{\rm UV})$~\cite{ZV}.
All of them are listed, for example,  in Ref.~\cite{ML-2loop}.

The contribution from the remaining diagrams shown
in Fig. 2 (b)-(e), which have gluon couplings to the eikonal lines,
is expressed as $\Delta$.
Their values are expressed in terms of one-loop scalar integrals:
\begin{eqnarray}
\Lambda^{\rm LC}&=&
i\;
\frac{2\pi\alpha_s\mu^{2\epsilon}}{3}
\bigg[9\left(7+\frac{ 1}{N}     \right)\;I_{AB}
     + \left(N+\frac{18}{N} - 67\right)\;I_{AD}
     -p^2 I_{AAD}
\nonumber\\
&&\quad\quad\quad\quad\quad
     +2\;p\cdot n\; \left(9I_{ACD} + I_{BCD}-36I_{ABC}\right)
\bigg],
\label{V:LC}
\\
\Lambda^{\rm F}&=&
i\;
\frac{2\pi\alpha_s\mu^{2\epsilon}}{3}
\bigg[ 9\left(1+\frac{ 1}{N}     \right)I_{AB}
     +  \left(N+\frac{18}{N} - 47\right)I_{AD}
     -p^2 I_{AAD}
\bigg],
\label{V:F}
\\
\Delta^{\rm LC}&=&0,
\\
\Delta^{\rm F}&=&
i\;
12\;{\pi\alpha_s\mu^{2\epsilon}}
\bigg[I_{AB}-2I_{AD}
      + p\cdot n\left(I_{ACD}+I_{BCD}\right) \bigg].
\label{DELTA:F}
\end{eqnarray}
The explicit value of the vertex correction factor $\Lambda^{\rm F}$ in the
Feynman gauge shown in Eq.~(\ref{V:F}) agrees with the result in 
Ref.~\cite{Braaten:2000pc}.
The UV dependence of the vertex correction factor $\Lambda^{\rm LC}$ 
in the light-cone gauge shown in Eq.~(\ref{V:LC}) agrees with the result using 
the ML prescription in Refs.~\cite{ZV,ML-2loop} where only the UV contribution
is given :
$\Lambda^{\rm LC}_Q=-\delta Z^{\rm LC}_Q=-\alpha_s/(3\pi\epsilon_{\rm UV})$.
The integral $I_{AAD}$ has a Coulomb singularity  
as well as a logarithmic IR divergence due to the exchange of a gluon between 
the on-shell heavy quark and anti-quark. 
Dimensional regularization puts power infrared divergence like the
Coulomb singularity to zero, so only the logarithmic IR divergence remains in
the integral $I_{AAD}$. Then the integral is effectively expressed by $I_{ADD}$
via the equation $I_{ADD}=(N-4)I_{AAD}$.
It is important to notice that various correction factors in 
the Feynman and the light-cone gauge involve different combinations of the
same scalar integrals. 
Straight-forward sums for both gauges produce a common result
\begin{eqnarray}
d_T^{\rm (virtual)}(z) = d_T^{(\rm LO)}(z)
\frac{4\pi\alpha_s}{3}
\;{\rm Re}\bigg\{i\left[
            -\left(7N-\frac{18}{N}+51\right) I_{AD}
      +  6n_f\left(1 -\frac{ 1}{N}   \right) I_{AB}
\right.
\nonumber\\
\left.
     + 18\;p\cdot n\left(I_{ACD}+I_{BCD}\right)
     + p^2\left(8I_{ADD}-I_{AAD}\right)
\right]
\bigg\}.
\label{dT-sum}
\end{eqnarray}
Thus the non-vanishing contributions from the gluon
coupling to the eikonal line in the Feynman gauge, $\Delta^{\rm F}$, is 
simply distributed to other correction factors in the light-cone gauge 
via additional gluon propagator terms.

Since gauge invariance holds for both the virtual and the real-gluon corrections
separately, the equality of the virtual corrections in the Feynman and the 
light-cone gauge is a consequence of gauge invariance.
As we commented previously, the light-cone dependent integrals
in the Feynman gauge result have no ambiguities form spurious poles.
On the other hand, we have not fixed the sign of the $i\epsilon$ in the 
spurious pole of the integrals which are obtained in the light-cone gauge.
Since we have found exact agreement between the two results in the two gauges,
we may simply use the values obtained from the Feynman gauge calculation.
Note that the integral $I_{ABC}$ disappears in Eq.~(\ref{dT-sum}), so the only 
light-cone dependent integrals that survive are  $I_{ACD}$ and $I_{BCD}$.
The values for these integrals are independent of the sign of the $i\epsilon$
in the definition of $C$ in Eq.~(\ref{eq:C}).  
The expansion of Eq.~(\ref{dT-sum}) 
in $\epsilon$ reproduces the result of Braaten and Lee~\cite{Braaten:2000pc}:
\begin{eqnarray}
&&d_T^{\rm (virtual)}(z) = d_T^{(\rm LO)}(z)\;
\frac{\alpha_s}{\pi}  \left( \frac{\pi \mu^2 }{m_Q^2} \right)^\epsilon
\nonumber\\
&&
\quad\quad
 \times
\left[  \frac{3(1-\epsilon)}{2}\frac{\Gamma(1+\epsilon)}{ 
         \epsilon_{\rm UV} \epsilon_{\rm IR}} 
	+ \beta_0 \frac{\Gamma(1+\epsilon)}{\epsilon_{\rm UV}}
	+ \frac{177-10n_f }{ 18} - \frac{\pi^2}{ 2} 
	+ 8 \ln 2 + 6 \ln^2 2  \right],
\label{d-virtual}
\end{eqnarray}
where $\beta_0=(33-2n_f)/6$.

\begin{figure}
\includegraphics[width=11.5cm]{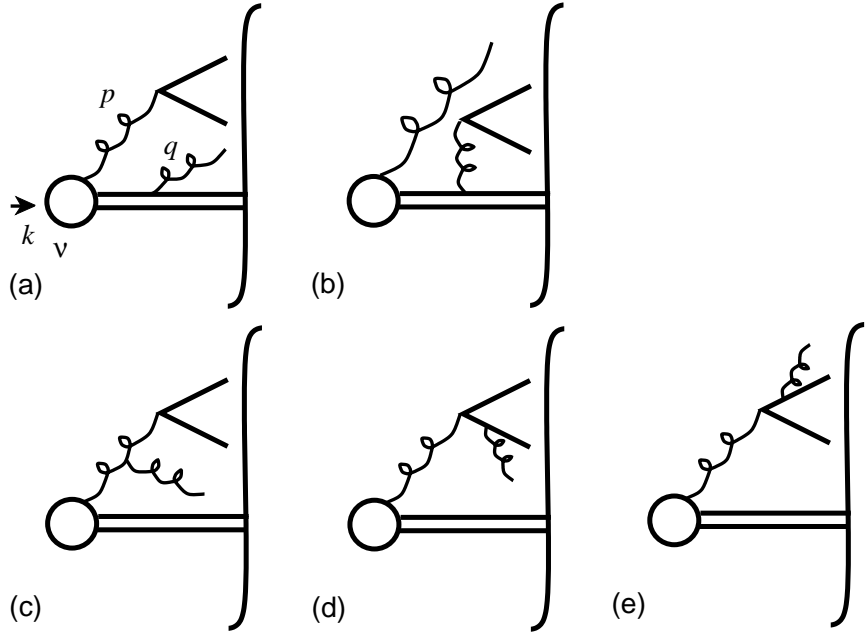}
\caption{\label{3body}
The Feynman diagrams of order $\alpha_s^2$ for $g \to Q\overline{Q}$
with $Q\overline{Q}g$ final states.  There are a total of 25 diagrams,
but only the left halves of the diagrams are shown.
}
\end{figure}

The Feynman diagrams for the real-gluon corrections to the fragmentation
function for $g\to Q\overline{Q}$ can also be calculated in both 
gauges. We draw the 5 left-half diagrams only, which must be multiplied by
their complex conjugates to give a total of 25 diagrams.
The real-gluon correction is a tree-level calculation.
Therefore, there is  no spurious pole problem.
In the Feynman gauge, all 25 diagrams contribute, while only
9 diagrams in the light-cone gauge. In the latter gauge,
diagrams \ref{3body}(a) and \ref{3body}(b) vanish.
The real-gluon correction contribution 
is also gauge invariant.
Employing either gauge,
we reproduce the real-correction contribution given in 
Ref~\cite{Braaten:2000pc}
before the phase-space integral is performed:
\begin{eqnarray}
d_T^{\rm  (real)}(z)&=&
\frac{\pi\alpha_s\mu^{2\epsilon}}
{8N(N-1)m_Q^3}
\times
\frac{3\alpha_s}{\pi\Gamma(1-\epsilon)}
\left(
\frac{\pi \mu^2}{m_Q^2}
\right)^\epsilon
\left(1-\frac{1}{z(1-z)}\right)^2
\int_{(1-z)/z}^\infty
dx
\frac{t^{1-\epsilon}}{x^2}\;,
\label{eq:epsIR}
\\
d_L^{\rm  (real)}(z)&=&
\frac{\pi\alpha_s\mu^{2\epsilon}}
{8Nm_Q^3}
\times
\frac{3\alpha_s}{\pi\Gamma(1-\epsilon)}
\left(
\frac{\pi \mu^2}{m_Q^2}
\right)^\epsilon
\left(\frac{1-z}{z}\right)^2
\int_{(1-z)/z}^\infty
dx
\frac{t^{-\epsilon}}{x^2}\;,
\end{eqnarray}
where $t=(1-z)(zx+z-1)$, $x=2q\cdot p/p^2$, 
$q$ is the final-state gluon momentum, and $p$ is the 
$Q\overline{Q}$ momentum. 
The final results for the real-gluon correction contribution of Braaten and 
Lee~\cite{Braaten:2000pc} are straight-forwardly reproduced:
\begin{eqnarray}
d_T^{\rm (real)}(z)
&=&
\frac{\pi\alpha_s\mu^{2\epsilon}}
{8N(N-1)m_Q^3}
\times
\frac{\alpha_s}{\pi}
\;
\left(\frac{\pi\mu^2}{m_Q^2}\right)^\epsilon \Gamma(1+\epsilon)
\nonumber\\
&&\times
\;\bigg[
-\frac{3(1-\epsilon)}{2\epsilon_{\rm UV}\epsilon_{\rm IR}}
\delta(1-z)
+\frac{3(1-\epsilon)}{\epsilon_{\rm UV}}
\left(
\frac{z}{(1-z)_+}+\frac{1-z}{z}+z(1-z)
\right)
\nonumber\\
&&
\quad\;\;\;
  -\frac{6}{z}\left(\frac{\ln(1-z)}{1-z}\right)_+
+6(2-z+z^2)\ln(1-z)
\bigg]\;,
\label{d-real}
\\
d_L^{\rm (real)}(z)
&=&
\frac{\pi\alpha_s}
{8Nm_Q^3}
\times
\frac{3\alpha_s}{\pi}
\frac{1-z}{z}\;.
\label{dL-real}
\end{eqnarray}

The infrared divergence cancels after summing the real and virtual 
correction contributions shown in Eqs.~(\ref{d-virtual}) and (\ref{d-real}).
Employing the $\overline{\rm MS}$ scheme, $\alpha_s$ and the operator are
renormalized as in Ref.~\cite{Braaten:2000pc}. After renormalization, 
the final answers for $d_T(z)$ and $d_L(z)$ of 
Braaten and Lee~\cite{Braaten:2000pc} are reproduced:
\begin{eqnarray}
d_T(z,\mu)
=
\frac{\pi\alpha_s(\mu)}{48m_Q^3}
\;
&\bigg\{&
\;
\delta(1-z)+\frac{\alpha_s(\mu)}{\pi}
\bigg[
A(\mu)\delta(1-z)
+\left(
 \ln\frac{\mu}{2m_Q}-\frac{1}{2}
 \right)P_{gg}(z)
\nonumber\\
&&+6(2-z+z^2)\ln(1-z)
  -\frac{6}{z}\left(\frac{\ln(1-z)}{1-z}\right)_+
\;
\bigg]
\bigg\}
\;,
\label{dT-final}
\end{eqnarray}
where the coefficient $A(\mu)$ is
\begin{eqnarray}
A(\mu)=
\beta_0\left(\ln\frac{\mu}{2m_Q}+\frac{13}{6}\right)
+\frac{2}{3}-\frac{\pi^2}{2}+8\ln 2+6\ln^2 2
\;,
\end{eqnarray}
and $P_{gg}(y)$ is the gluon splitting function:
\begin{eqnarray}
P_{gg}(z)=
6\left[
\frac{z}{(1-z)_+}+\frac{1-z}{z}+z(1-z)+\frac{\beta_0}{6}\;\delta(1-z)
\right]\;.
\end{eqnarray}
The transverse term $d_T(z)$ in Eq.~(\ref{dT-final}) still disagrees 
with that of Ref.~\cite{Ma-2}.
Our final answer for the longitudinal fragmentation function is 
obtained by setting $\epsilon \to 0$ in Eq.~(\ref{dL-real}):
\begin{eqnarray}
d_L(z,\mu)
=\frac{\alpha_s^2(\mu)}{8m_Q^3}\;\frac{1-z}{z} \,.
\label{dL-final}
\end{eqnarray}
The longitudinal term, $d_L(z)$ agrees with that of 
Braaten and Lee~\cite{Braaten:2000pc}
as well as that of Beneke and Rothstein~\cite{Beneke:1995yb}.
The dependence on the spectroscopic state of the produced quarkonium 
of this fragmentation function can be found in Ref.~\cite{Braaten:2000pc}.

\section{discussion}
We have calculated the next-to-leading order correction to 
the color-octet $^3S_1$ term in the gluon fragmentation function in a 
light-cone gauge without using conventional prescriptions for the spurious
pole. The gauge-invariant definition of the fragmentation function 
of Collins and Soper allows us to fix the ambiguities from spurious poles 
in the light-cone gauge by comparing with the result obtained in
the Feynman gauge. Our result agrees with that of 
Braaten and Lee~\cite{Braaten:2000pc} which disagrees with that of 
Ref.~\cite{Ma-2}.
The light-cone gauge considerably simplifies the calculation procedure for 
both the real and the virtual corrections.
At least at one-loop level, the spurious pole problem can be resolved. 
This problem does not appear in the real corrections
because they come from tree-level diagrams, but it does appear in the virtual 
corrections.  
We reduced the virtual correction in the color-octet $^3S_1$ 
fragmentation function in the light-cone gauge to a linear combination
of scalar integrals. After naive cancellations among the scalar integrals,
ignoring the ambiguity from spurious poles,
the correction reduces to scalar integrals that are independent of the sign
of $i\epsilon$ in the denominator $k\cdot n+i\epsilon$. 
As a byproduct, the renormalization constants in the
light-cone gauge were obtained at one-loop level.  Their UV dependencies agree 
with the previous calculations within the ML prescription. 
They might be useful for 
other calculations, such as the next-to-leading order corrections to other 
fragmentation functions.

\begin{acknowledgments}
I would like to acknowlege that this work is benefitted from a number of
helpful comments from George Leibbrandt.
I thank Eric Braaten and Gustav Kramer
for their valuable comments.
This work is supported by a Korea Research Foundation 
Grant(KRF-2004-015-C00092).
\end{acknowledgments}

\end{document}